\documentclass[%
 aip,
 amsmath,amssymb,
 reprint,%
]{revtex4-1}

\usepackage{graphicx}
\usepackage{dcolumn}
\usepackage{bm}
\usepackage{xcolor}

\usepackage[utf8]{inputenc}
\usepackage[T1]{fontenc}
\usepackage{etoolbox}
\usepackage{physics} 
\usepackage{siunitx}
\usepackage{newtxtext}
\usepackage[varvw]{newtxmath}
\usepackage[version=3]{mhchem}

\graphicspath{{figures/}} 

\makeatletter
\def\@email#1#2{%
 \endgroup
 \patchcmd{\titleblock@produce}
  {\frontmatter@RRAPformat}
  {\frontmatter@RRAPformat{\produce@RRAP{*#1\href{mailto:#2}{#2}}}\frontmatter@RRAPformat}
  {}{}
}%
\makeatother
\begin{document}

\preprint{AIP/123-QED}

\title[]{Self-consistent formation and steady-state characterisation of trapped high energy electron clouds in the presence of a neutral gas background}
\author{G. Le Bars}
\email{guillaume.lebars@epfl.ch.}

\author{J.-Ph. Hogge}%
\author{J. Loizu}
\author{S. Alberti}
\author{F. Romano}

\affiliation{ 
Ecole Polytechnique Fédérale de Lausanne (EPFL), Swiss Plasma Center (SPC), CH-1015 Lausanne, Switzerland 
}%
\author{A. Cerfon}
\affiliation{ 
Courant Institute of Mathematical Sciences, New York University, New York, NY 10012 USA 
}%
\date{\today}

\begin{abstract}
This study considers the self-consistent formation and dynamics of electron clouds interacting with a background neutral gas through elastic and inelastic (ionisation) collisions in coaxial geometries similar to gyrotron electron guns. These clouds remain axially trapped as the result of crossed magnetic field lines and electric equipotential lines creating potential wells similar to those used in Penning traps. Contrary to standard Penning traps, in this study we consider a strong externally applied radial electric field which is of the same order as that of the space-charge field. In particular, the combination of coaxial geometry, strong radial electric fields and electron collisions with the residual neutral gas (RNG) present in the chamber induce non-negligible radial particle transport and ionisation. In this paper, the dynamics of the cloud density and currents resulting from electron-neutral collisions are studied using a 2D3V particle-in-cell code. Simulation results and parametric scans are hereby  presented. Finally, a fluid model is derived to explain and predict the cloud peak density and peak radial current depending on the externally applied electric and magnetic fields, and on the RNG pressure. 
\end{abstract}

\maketitle

\section{Introduction}
\label{sec:introduction}

Nonneutral plasmas are relevant to many fields in physics and engineering, for example elementary particle physics~\cite{vogel_particle_2018}, particle accelerators~\cite{davidson_physics_2001}, or high power microwave sources such as gyrotrons~\cite{chu_electron_2004}. They present confinement properties and types of instabilities that make them fundamentally different from quasi-neutral plasmas~\cite{davidson_physics_2001}. Nonneutral plasmas have been studied since the 1960s in various configurations, the most common being the Penning-Malmberg trap: a cloud of single sign charged particles is trapped in a cylindrical chamber by a strong axial magnetic field and two end electrodes imposing a confining axial electric field~\cite{dubin_trapped_1999}. This configuration is predominantly used to study the behaviour of elementary particles at low and extremely low temperatures~\cite{dubin_trapped_1999}.

In Penning-Malmberg traps, the equilibrium and dynamics of charged particle clouds (electrons or ions) are dominated by long-range collective effects associated with the electric field self-generated by the plasma. The total trapped charge is controlled externally through the axial magnetic field, the end electrodes voltage, and the external particle source~\cite{vogel_particle_2018}. The level of vacuum inside the device is such that ionisation processes between the confined charges and the background neutrals can be neglected. To ensure that the equilibrium state exists over very long time-scales~\cite{PhysRevLett.44.654}, many dedicated theoretical and experimental studies have been carried out to avoid disruptive instabilities such as the azimuthal diocotron instability~\cite{doi:10.1063/1.5018425,doi:10.1063/1.4875341,Kartashov2010,Petri2009,doi:10.1063/1.2040177,UCLA,doi:10.1063/1.1302121,Schuldt:1995pb,PhysRevLett.64.645,ricketson2016sparse,Muralikrishnan2021} and, to a lesser extent, axial resonant space-charge effects~\cite{vogel_particle_2018}. The confining electric potential along the magnetic field lines, which is imposed externally, is generally of the order of tens of \SI{}{\volt} since the plasma temperature is of the order of a few \SI{}{\electronvolt}~\cite{dyavappa_dependence_2017}. The cloud dynamics is characterised by a net rotation velocity that is generally non-relativistic and is given by the $\vec{E}\cross\vec{B}$ drift, with the radial component of the electric field, $E_r$, determined by the cloud space-charge. This rotational velocity provides an inward-pointing $\vec{v}\cross\vec{B}$ force that balances the radially outward electric force.

The study of nonneutral plasmas is also relevant to the development of gyrotron electron guns, where secondary electron clouds (i.e. not belonging to the main electron beam) can remain locally trapped, and accumulate in effective potential wells due to the local electric and magnetic field topology~\cite{pagonakis_gun_2009,pagonakis_electron_2016}. The trapped electrons, if released by instabilities or shielding of the wells by space-charge effects, can lead to detrimental and even damaging currents that prevent nominal gyrotron operation. These potential well traps are very similar to the ones used in Penning-Malmberg traps. However, in the case of the gyrotron gun, the magnetic field is non-uniform and the electric field has a strong, externally imposed, radial component, with only a fraction of the total electric field given by the trapped electron cloud space-charge itself. Due to the resulting strong azimuthal $\vec{E}\cross \vec{B}$ drift, the trapped electrons have sufficient kinetic energy to ionise the residual neutral gas (RNG) in the gyrotron vacuum chamber. Such ionisation phenomena, combined with the existence of trapping potential wells, have been linked to the observed detrimental currents in gyrotron guns~\cite{pagonakis_electron_2016}. To avoid this problem, a design criterion has been defined which imposes the absence of any vacuum potential well in the electron gun by a careful modification of the electrodes shapes~\cite{pagonakis_gun_2009}. However, this process becomes highly demanding from an engineering point of view, and relaxed design criteria are highly desirable. This work is particularly relevant to the research effort in magnetically confined fusion plasmas, as gyrotrons are foreseen to play a major role in heating, current drive, and instability control of fusion plasmas~\cite{Alberti:2007fk} in tokamaks, such as ITER~\cite{Darbos:2016ee}, DDT~\cite{AMBROSINO2021112330}, or DEMO~\cite{GARAVAGLIA20181173}, and in stellarators such as Wendelstein 7-X~\cite{laqua_overview_2019}.

As of today, there has been no study, to the authors knowledge, of electron trapping in chambers with azimuthal symmetry that are subject to strong external radial electric fields, {\it and} in which electron-neutral collisions dominate the sources and sinks of electrons. Collisions between trapped particles and RNG in the vacuum chamber lead to radial transport. Such phenomena have been studied using a kinetic model~\cite{douglas_transport_1978} or a fluid model~\cite{davidson_nonlinear_1996,davidson_expansion_1996} but without considering ionisation as a source of electrons. Studies including ionisation effects have been done using a fluid model considering cold~\cite{javakhishvili_nonlinear_1999} or hot isothermal plasmas~\cite{kervalishvili_collisional_2000} but neglecting inertial effects. Furthermore, none of these studies considered non-uniform magnetic fields or external radial electric fields.

This study presents fully-kinetic numerical simulations that simultaneously take into account electron-neutral elastic and inelastic (ionisation) collisions, strong externally imposed electric fields, as well as axial and radial non-uniformities typical of gyrotron guns (geometry, magnetic field, external electric field). To explain the simulation results, an analytical fluid model is derived that highlights the parametric dependence of peak density and leaking currents associated with the formation of electron clouds.

A 2D3V particle-in-cell (PIC) code, assuming azimuthal symmetry, has been developed and used to characterise the self-consistent electron cloud formation in a gyrotron electron gun due to RNG ionisation. Contrary to the traditional Penning-Malmberg trap, the geometry of interest has a coaxial configuration, and the trapped electron cloud presents a hollow density profile. This is a significant difference compared to the filled cylindrical column stored in Penning-Malmberg traps, as it has been shown that annular nonneutral plasma clouds are highly susceptible to the diocotron instability~\cite{PhysRevLett.64.645,davidson_physics_2001}. However, this instability is not considered in the present study.

The article is divided into the following sections: section~\ref{sec:methods} describes the numerical method and the geometry of interest, while section~\ref{sec:results} presents the main findings from individual PIC simulations, specifically the type of losses, the dominant forces in each direction, and shows the results of parametric scans highlighting the dependencies of the cloud peak density and current on the key parameters of the system. To explain these dependencies, a fluid-Poisson model is derived and compared to the numerical results in section~\ref{sec:AnModel}. A summary of the simulation results and density and current dependencies on external parameters, and related discussions make up the last section of the paper, section~\ref{sec:discussion}.

\section{Methods}
\label{sec:methods}

\begin{figure*}
  \includegraphics[width=1\textwidth]{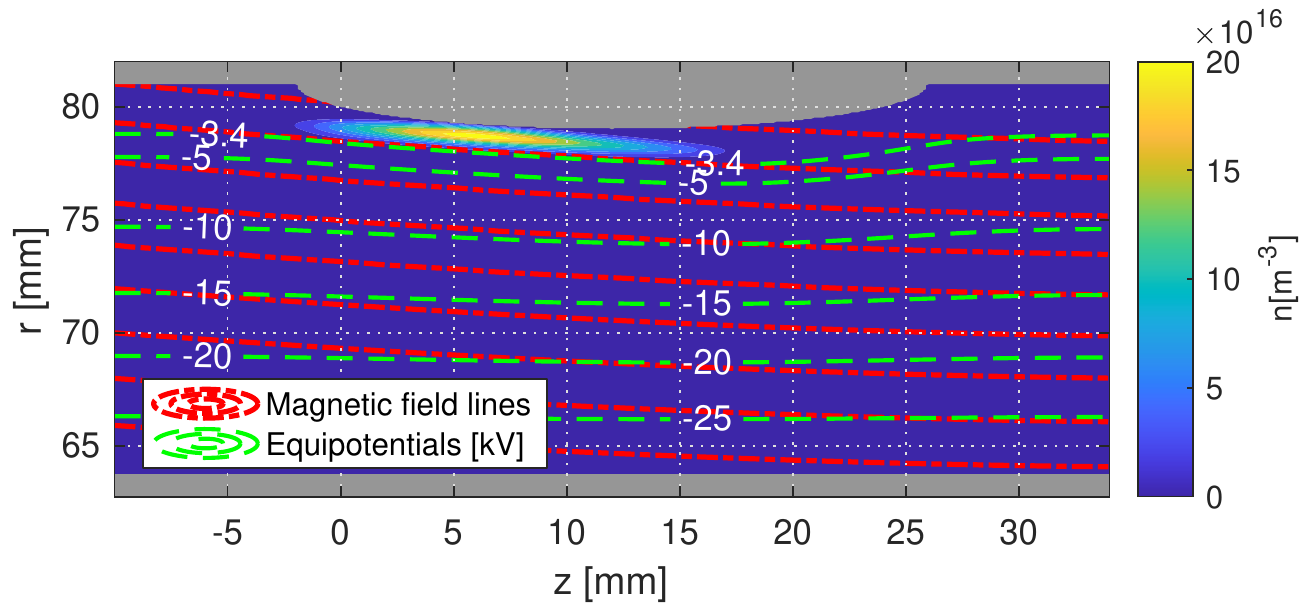}
  \caption{\label{fig:tes30kv_ince2Fields} Geometry and simulation domain. The top and bottom grey parts represent the metallic boundary conditions with an externally imposed bias of $\Delta\phi = \SI{30}{\kilo\volt}$. The peak magnetic field amplitude in this domain, here in the lower right corner, is $B_{max}=\SI{0.28}{\tesla}$. The cloud density and self-consistent electric equipotential lines are represented at a time when the number of trapped electrons is maximum.}
\end{figure*}

\subsection{Numerical model}
An electrostatic 2D3V, PIC code has been developed that solves the collisional Vlasov-Poisson equation for the electron distribution function $f_e(\vec{r},\vec{v},t)$ and the electric potential $\phi(\vec{r},t)$. This codes uses a cylindrical coordinate system $(r,\theta,z)$ and assumes azimuthal symmetry $(\partial_\theta=0)$. The Poisson equation is solved using a finite element method based on weighted extended b-splines of any order~\cite{hollig_finite_2003,hollig_weighted_2001}. This method allows the definition of Dirichlet boundary conditions on curved surfaces, while keeping a rectangular grid for the finite element method and allowing flexibility in the choice of geometries. With such numerical method, the non-trivial geometry of the electron gun can be easily approximated in the code without needing an optimized, often complex, mesh adapted to the particular geometry, as with traditional finite element methods based on triangular cells. Indeed, this technique improves the flexibility and rapidity to explore different geometries. The use of a rectangular grid also simplifies the parallelisation of the code using both multiple nodes (MPI) and multiple cores in each node (openMP). The Vlasov equation is solved for $f_e$ using the standard PIC method by sampling this distribution function with macro-particles and by calculating the trajectory of each macro-particle using the Boris algorithm~\cite{birdsall_plasma_2017}. The particle boundary conditions are perfectly absorbing, and the electric potential boundary conditions are a mix of Dirichlet on the metallic parts, and Neumann on the open boundaries. This is illustrated in Fig.~\ref{fig:tes30kv_ince2Fields}, where the grey parts, representing metallic boundaries, are set at a fixed potential, while Neumann boundary conditions are imposed at $z_L=\SI{-10}{mm}$ and $z_R=\SI{34}{mm}$ in the blue region, $\partial_z\phi(z=z_L)=\partial_z\phi(z=z_R)=0$.

To initiate the self-consistent cloud formation, a volumetric seed source of electrons is used as a low intensity source spanning the full simulation space. This creates electrons according to a uniform distribution function in space and a Maxwellian distribution function in velocity. This mimics the background, low-density, free-electrons present in the electron gun region due to field electron emission of the metallic surfaces, or due to ionisation of the RNG by natural background radiation. The source term is set such that, once a cloud is formed, the seed source is negligible compared to the ionisation source.

An electron-neutral collision module has been implemented using a Monte Carlo approach~\cite{birdsall_particle--cell_1991,sengupta_influence_2016}. This module simulates elastic and ionisation collisions of the electrons with the RNG in the vacuum chamber. It assumes that the neutral density and temperature are constant in time and uniform, and that only one type of gas is present. The nature and density of the gas are input parameters set at the beginning of the simulation. Furthermore, as the ion Larmor radius is typically bigger than the radial dimension of the chamber, the ions created by ionisation are rapidly lost and are not simulated. Concerning elastic collisions, the cross-section for momentum exchange is extracted from the LXCat library~\cite{Biagicross,BSR2021}, and the scattering angle, which is dependent on the incoming electron kinetic energy, is calculated using a random number generator and an analytical differential cross-section~\cite{okhrimovskyy_electron_2002}. For the inelastic ionisation collisions, the total ionisation cross-section is also extracted from the LXCat library~\cite{Biagicross,BSR2021}. At the moment of collision, the incoming electron kinetic energy is reduced by the binding energy of the freed electron, and is distributed unevenly between the incoming and freed electron. The ratio of kinetic energy between incoming and freed electrons is calculated using an energy differential cross-section based on experimental measurements~\cite{opal_measurements_1971}. The freed and incoming electron undergo a scattering event, and the scattering angles are calculated using the same analytical differential cross-section as for elastic collisions~\cite{okhrimovskyy_electron_2002}. In essence, the elastic collisions act as a drag term in the force balance equation for the electrons, while the ionising collisions act as a source of electrons in the continuity equation, and as an effective drag in the force balance equation. Details on the numerical model will be presented in a future publication.

\subsection{Geometry and numerical parameters}

\begin{figure}
\centering
  \includegraphics[width=0.45\textwidth]{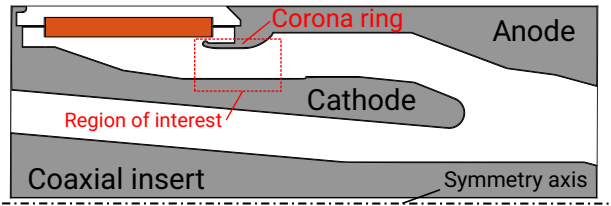}
  \caption{\label{fig:gt170_geom} Cut-view of the electron gun used in the first prototype of the \SI{170}{\giga\hertz} \SI{2}{\mega\watt} coaxial gyrotron designed for ITER. This configuration is used as a model for the geometry considered in this study, see Fig.~\ref{fig:tes30kv_ince2Fields}. This representation assumes azimuthal symmetry. Grey indicates a metallic component, orange indicates an insulator, and white represents vacuum. }
\end{figure}

\begin{figure}
  \includegraphics[width=0.48\textwidth]{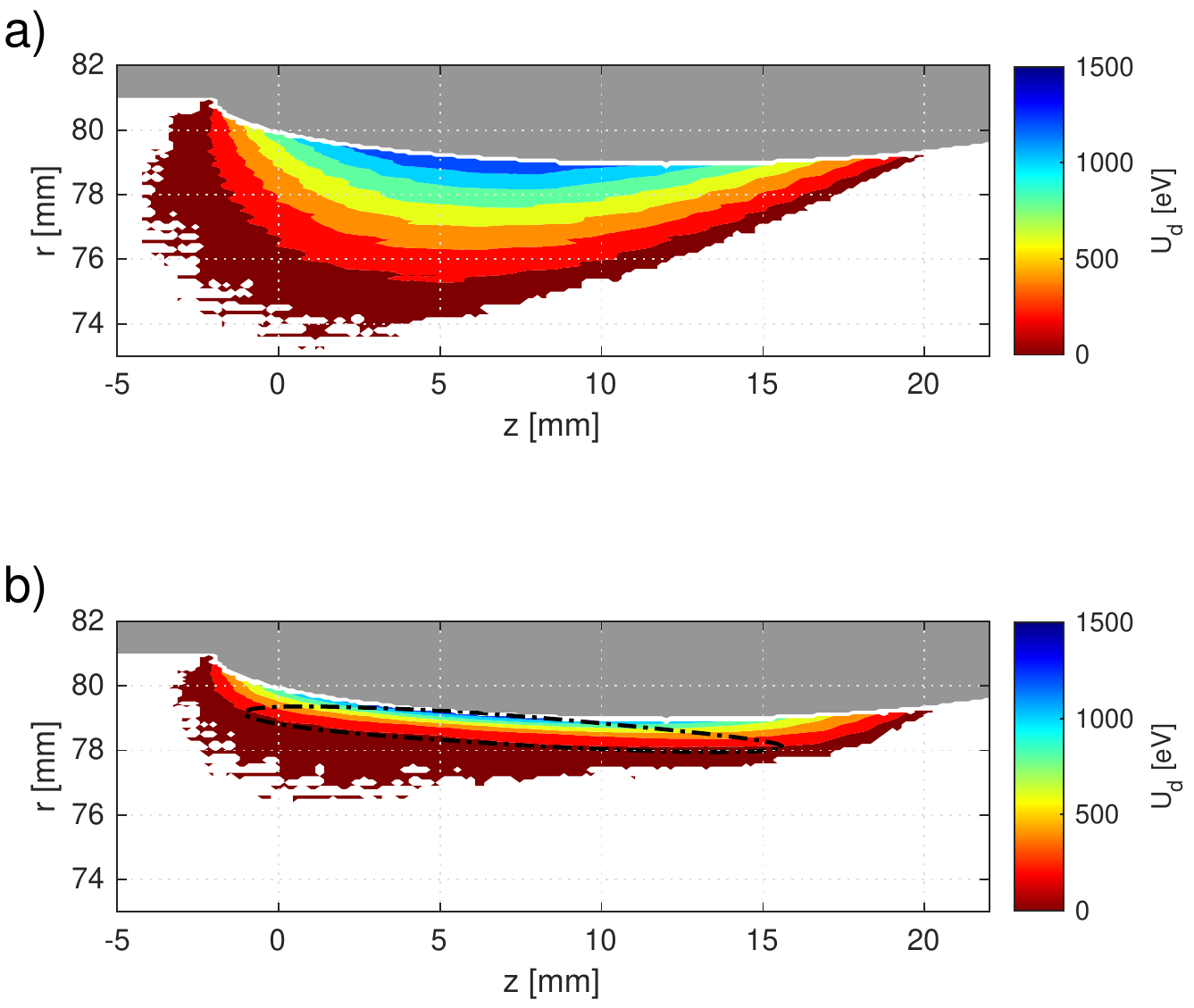}
  \caption{\label{fig:tes30kv_ince2Well} Potential well position and depth for the geometry of Fig.~\ref{fig:tes30kv_ince2Fields}. The plots are zoomed on the well region for readability. In a) the potential well is represented in the vacuum condition: no electron cloud is present. In b) the potential well is represented at peak electron density. The black dashed-dotted line represent the cloud edge defined as the positions where $n=0.2\;n_{e,max}$.}
\end{figure}

\begin{figure}
    \centering
    \includegraphics[width=0.29\textwidth]{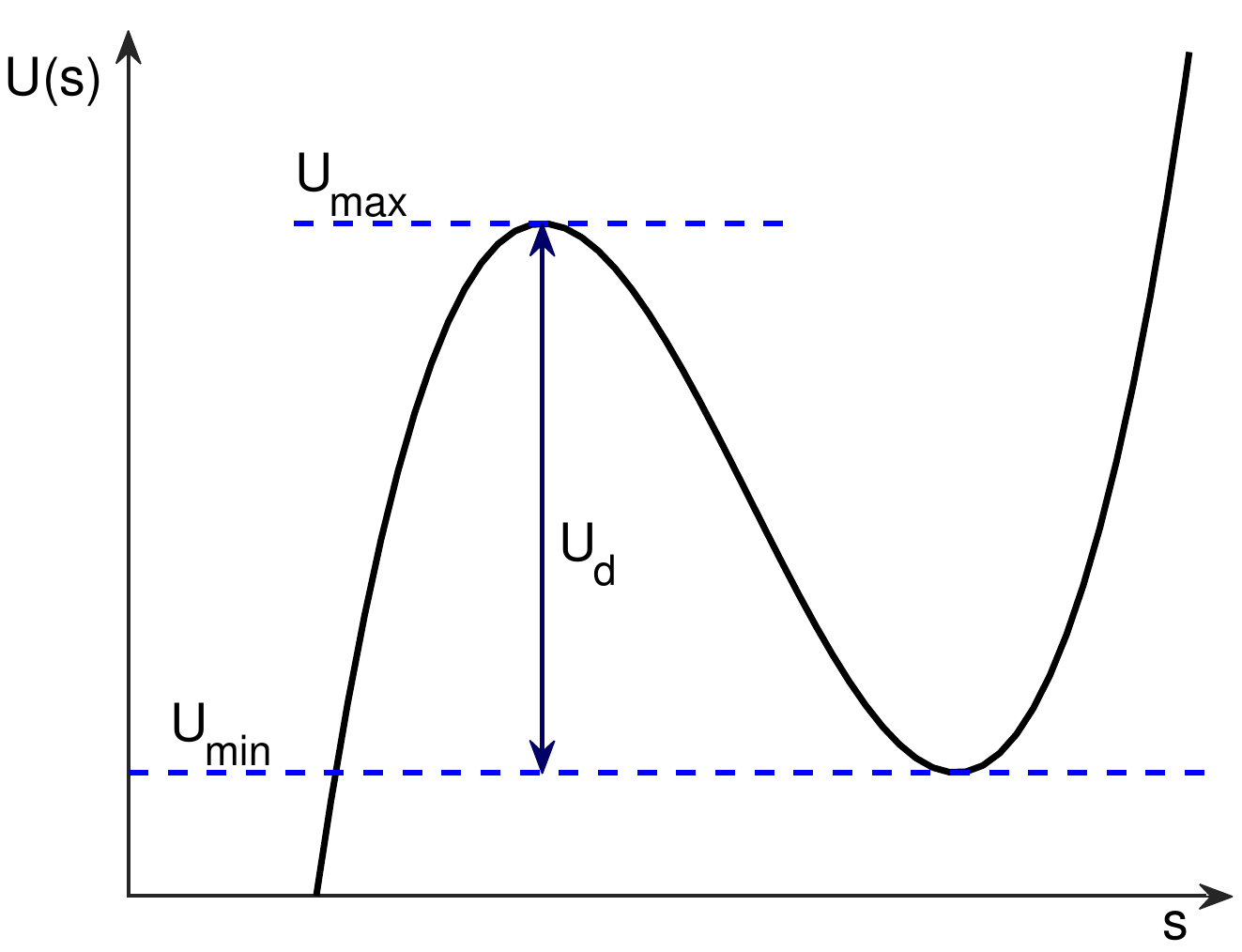}
    \caption{The potential well depth $U_d$ is defined for a specific magnetic field line as the difference between the local electric potential energy $U(s)$ and the highest local minimum $U_{min}$ along the magnetic field line coordinate $s$.}
    \label{fig:Uwell}
\end{figure}

The geometry used in this study, represented in Fig.~\ref{fig:tes30kv_ince2Fields}, is based on the electron gun geometry used in the first prototype~\cite{pagonakis_gun_2009} of the \SI{170}{\giga\hertz} \SI{2}{\mega\watt} coaxial gyrotron designed for ITER, see Fig.~\ref{fig:gt170_geom}. This particular Magnetron Injection Gun (MIG) was subject to voltage breakdown and detrimental leakage currents for specific magnetic field configurations, and had to be redesigned to allow nominal operation~\cite{pagonakis_gun_2009}. The configuration of this study focuses on the corona ring region of the original prototype gun, and uses the same magnetic field used for the prototype gyrotron with $B\sim\SI{0.28}{\tesla}$~\cite{pagonakis_magnetic_2017}. This region can be approximated by a coaxial configuration as represented in Fig.~\ref{fig:tes30kv_ince2Fields}, with a central conductor of radius $r_a=\SI{63.75}{\milli\meter}$, and an outer cylinder of radius $r_b=\SI{81}{\milli\meter}$. On the outer cylinder, a filled elliptic region centred at $r_0=\SI{81}{\milli\meter}$, $z_0=\SI{12}{\milli\meter}$, with "major axis" $\delta z=\SI{14}{\milli\meter}$, and "minor axis" $\delta r=\SI{3}{\milli\meter}$ is added. Between the central and the outer metallic parts, a bias $\Delta\phi$ is applied which, combined with the externally applied magnetic field $\vec{B}$, induces a strong azimuthal $\vec{E}\cross\vec{B}$ drift. Due to the upper elliptic region, the electric equipotential lines combined with the magnetic field lines topology lead to the formation of a potential well, see Fig.~\ref{fig:tes30kv_ince2Well}. In this study, a potential well for electron trapping is defined according to Pagonakis et al~\cite{pagonakis_electron_2016}. It is a region where the electric potential energy has a local maximum $U_{max}$ along a magnetic field line. The depth of this potential well is determined by the highest local minimum $U_{min}$ on both sides of $U_{max}$ along the magnetic field line. The maximum depth is $U_d=U_{max}-U_{min}$ as represented in Fig.~\ref{fig:Uwell}. In the selected case, the well depth spans the range $U_d=200-\SI{3600}{\electronvolt}$ for a bias $\Delta\phi$ ranging from $5$ to $\SI{90}{\kilo\volt}$. We remark that the values of $U_d$ result from a combination of the externally imposed electric field as well as that generated by the space-charge, and so it can only be known after the simulation is run. In Fig.~\ref{fig:tes30kv_ince2Well} both the externally imposed well in vacuum and the self-consistent well in the presence of the cloud are displayed. 

In the simulations, the RNG pressure is artificially increased to allow running for several collision times while resolving cyclotron motion. Indeed, in gyrotrons, due to the low RNG pressure of the order of $p_n=10^{-8}\ \SI{}{\milli\Bar}$, the collision frequencies are several orders of magnitude smaller than cyclotron or plasma frequencies. As the current code resolves the cyclotron motion, it is necessary to bring these three time-scales closer to each other. To this end, the neutral pressure is increased to $10^{-1}\SI{}{\milli\Bar}$. The relatively high pressure sets the collision time scales ($\tau_d\sim\SI{5e-9}{\second}$) closer to the cyclotronic ($\tau_{ce}\sim\SI{1e-10}{\second}$ at $B\sim\SI{0.28}{\tesla}$), and plasma time scales ($\tau_{pe}\sim\SI{2.5e-10}{\second}$ at $n\sim\SI{2e17}{\per\cubic\meter}$), while keeping sufficient time-scale separation, such that $\tau_d \gg \tau_{ce}$ and $\tau_d \gg \tau_{pe}$. Despite this time-scale "compression", the wall-clock time of a single simulation can be as large as $\sim~1$ day when running on 36 cpus. As shown later in section~\ref{sec:results}, the results of this study show that simple scaling laws exist for RNG pressure effects that support the choice of high pressures for numerical simulations. This also allows direct extrapolation of the results to arbitrarily low neutral pressures. In addition, due to the general availability of total and differential cross-section data, and the possibility of future verification against experimental data, the background gas considered in the simulations is \ce{Ne}. Nevertheless, the simulation results and the reduced model presented in this paper are general and, given the appropriate collision cross-sections, can be directly adapted to different kinds of gases such as \ce{H2} or \ce{H2O}, typically present in vacuum vessels~\cite{mukherjee_hydrogen_2021}.

\section{Results}
\label{sec:results}
\subsection{Time evolution}
\begin{figure}
    \includegraphics[width=0.48\textwidth]{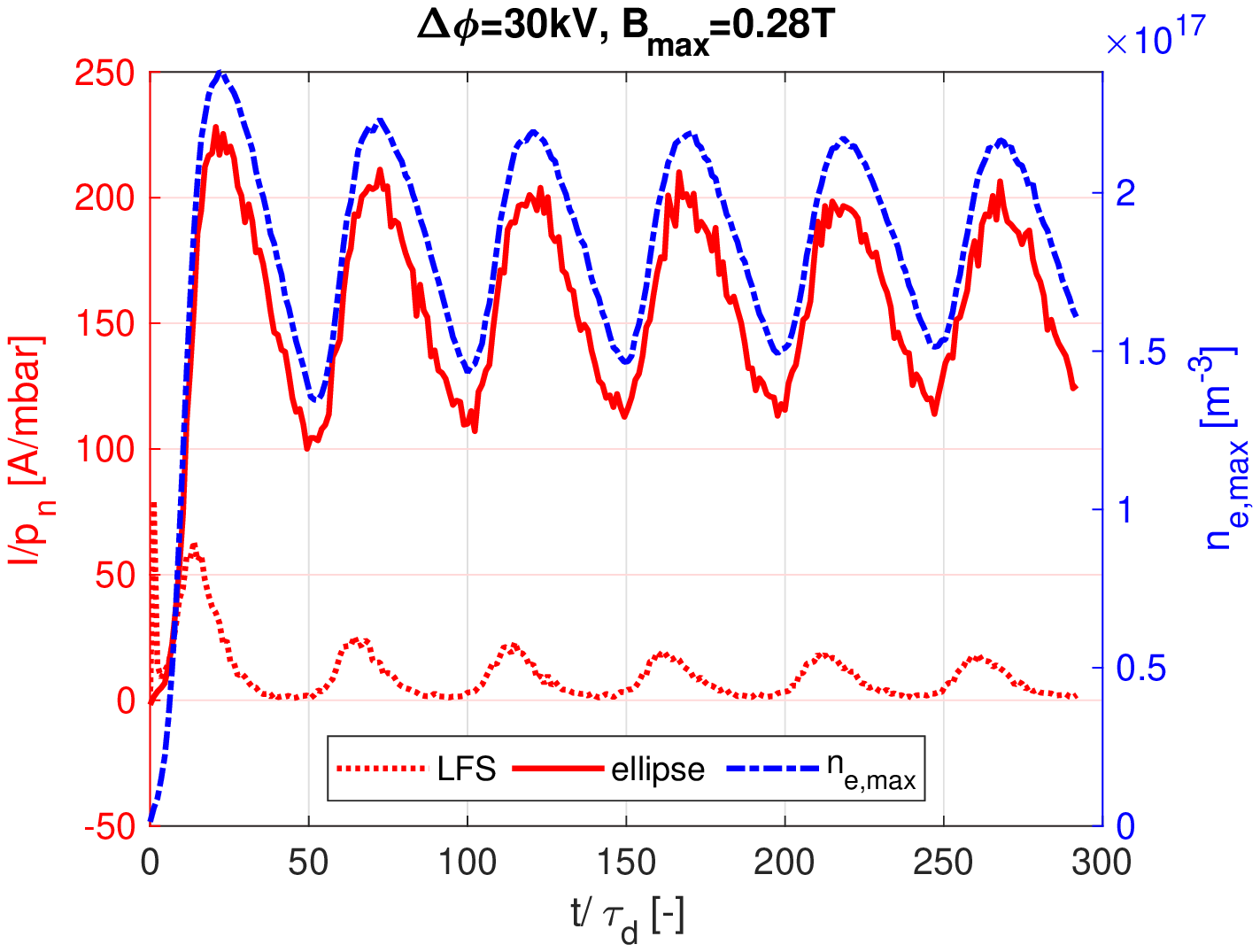}
    \caption{\label{fig:tes30kv_timeevol} Time evolution of the peak electron density in the simulation domain (blue) and total axial and radial currents (red). The currents are divided by the RNG pressure in \SI{}{\milli\Bar} and the time is normalised to the total collision characteristic time-scale for the momentum exchange. LFS (Low magnetic Field Side) is the axial current at $z_L=\SI{-10}{\milli\meter}$. "ellipse" is the total radial current collected on the elliptic metallic part. The current on the other boundaries are negligible throughout the simulation.}
\end{figure}

\begin{figure}
    \centering
    \includegraphics[width=0.48\textwidth]{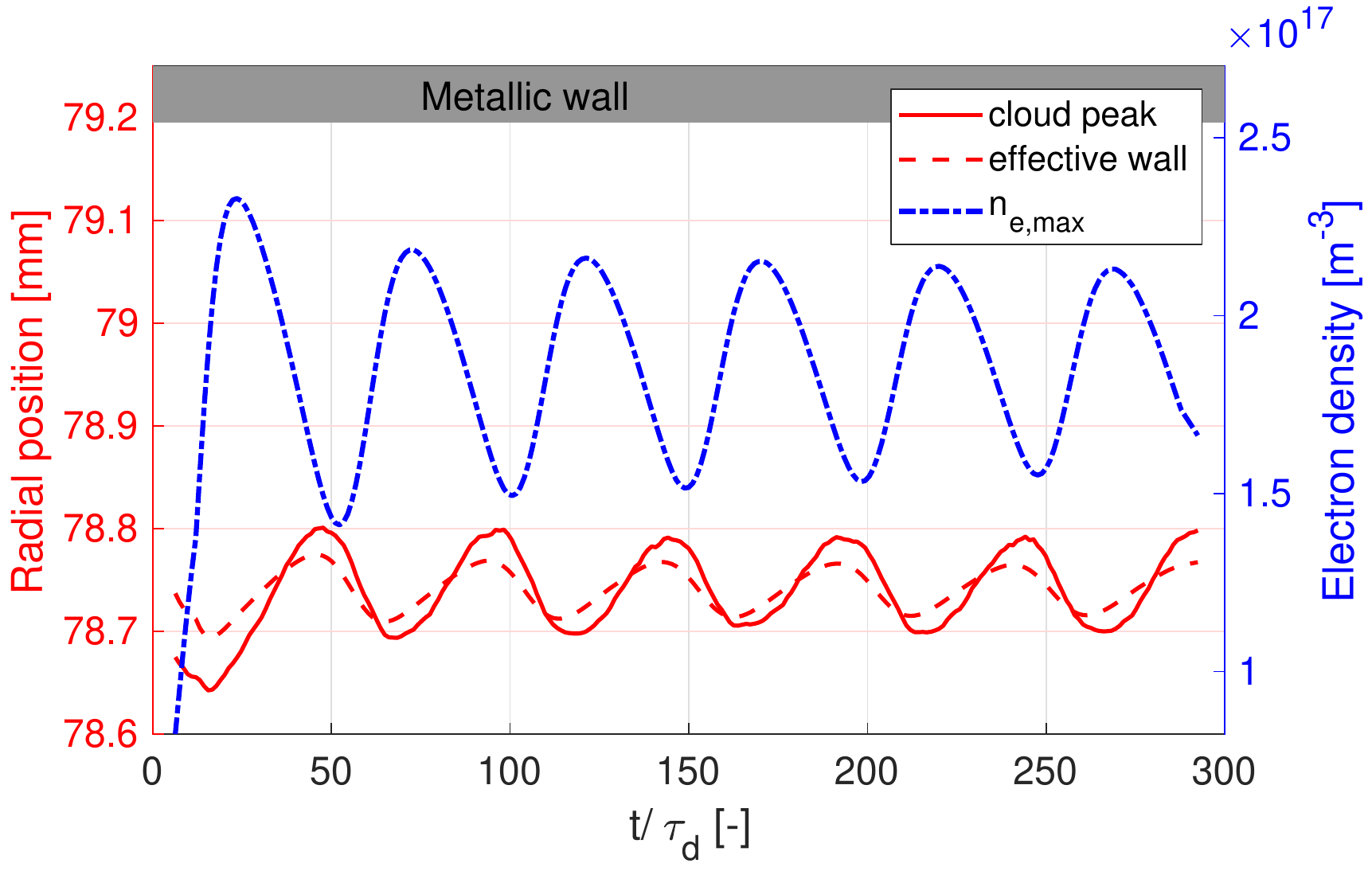}
    \caption{Time evolution of the instantaneous spatial cloud peak density radial position (solid red), and of the  effective wall radial position (dashed red) for the simulation shown in Fig.~\ref{fig:tes30kv_ince2Fields} and Fig.~\ref{fig:tes30kv_timeevol}. The effective wall is defined as $r_{wall}(z_{peak})-2\rho_L(r_{peak},z_{peak},t)$ with $r_{wall}(z_{peak})$ the upper metallic wall radial limit at the axial cloud peak position $z_{peak}$, and $\rho_L(r_{peak},z_{peak},t)$ the instantaneous Larmor radius at the cloud peak position $(r_{peak},z_{peak})$.}
    \label{fig:rpos}
\end{figure}

The simulations are initialised, in the presence of a \ce{Ne} gas background, with a homogeneous low density cloud following a Maxwellian distribution with $n_0=\SI{1e15}{\m^{-3}}$ and temperature $T_0=\SI{1}{\electronvolt}$ acting as a seed for the ionisation process. The electrons outside of the vacuum potential well, see Fig.~\ref{fig:tes30kv_ince2Well}, are rapidly lost axially. The remaining trapped electrons collide with the RNG, leading to the formation of a cloud. This cloud is located close to the elliptic region where the potential well is deepest, see Fig.~\ref{fig:tes30kv_ince2Fields}. The cloud density then slowly increases over time due to ionisation and trapping of the newly created electrons. As shown in Fig.~\ref{fig:tes30kv_timeevol}, as the cloud density increases, a radial current, and a comparatively smaller axial current establish, leading to charge losses at the boundaries of the simulation domain. During the cloud formation, the electrons drift radially because of the effective azimuthal drag caused by electron neutral collisions. This drift induces an outward-going radial motion of the cloud peak density, see Fig.~\ref{fig:rpos}. At the same time, the density increase causes an increase of the radial electric field amplitude. As the electron perpendicular velocity is strongly dependent on the $\vec{E}\cross\vec{B}$ drift, the increase in $E_r$ induces an expansion of the electron Larmor radius $\rho_L$, which eventually produces particle losses due to gyro-orbits intersecting the wall. It is thus useful to define an "effective wall" as a virtual surface distanced by two Larmor radii from the metallic wall. This effective wall defines the radial limit above which electrons can potentially hit the metallic boundary. As shown in Fig.~\ref{fig:rpos}, the combination of $\rho_L$ expansion and radial drift of the cloud peak density lead to a moment when the density peak radial position is above the effective wall position. This induces capture of electrons belonging to the cloud peak density by the metallic wall, causing an important radial loss. As the electron source is directly proportional to the electron density, the system is effectively subjected to a modulated source and a modulated sink, and gives rise to oscillations in the cloud density. This effect can be observed in Fig.~\ref{fig:tes30kv_timeevol}, where both the maximum cloud density and the boundary currents reach a peak after a few tens of $\tau_d$, when the losses start to dominate and oscillations in the peak density and radial current develop. In the absence of a steady electron seed source, the cloud is completely lost radially after several tens of $\tau_d$. However, the presence of a steady electron seed source can restart the cloud formation process and the cloud density oscillates between a minimum and a maximum value, in a periodic manner, in what could be called "cloud breathing", see Fig.~\ref{fig:tes30kv_timeevol}. It can also be observed from Fig.~\ref{fig:tes30kv_timeevol}, that the losses are dominantly radial, while axial confinement remains almost ideal.

\subsection{Parametric scans}
\begin{figure}
    \centering
    \includegraphics[width=0.45\textwidth]{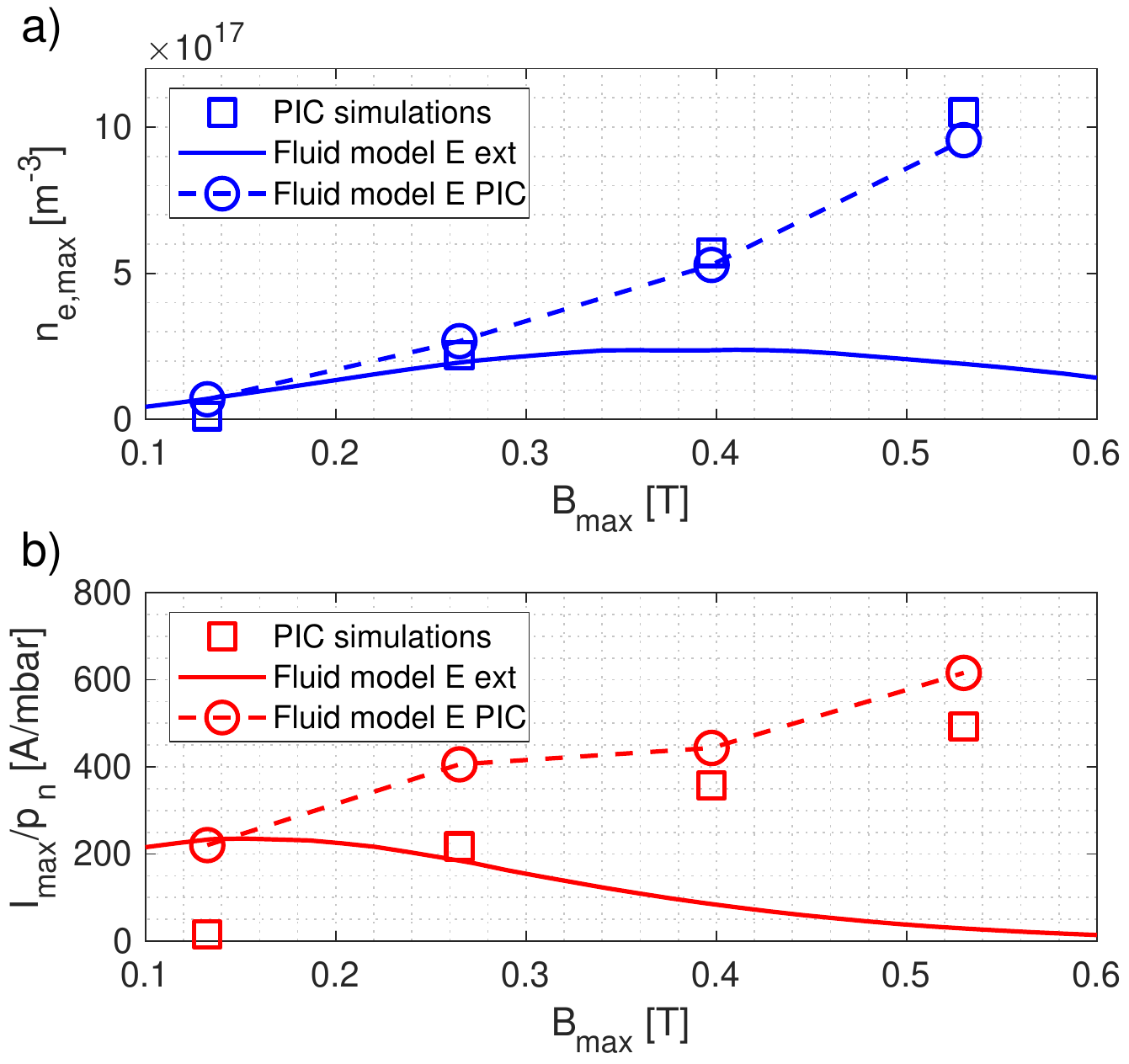}
    \caption{a) Evolution of the maximum electron density in the cloud as a function of the maximum magnitude of the magnetic field in the cloud region. b) evolution of the maximum radial current, scaled by the RNG pressure, as a function of the maximum magnitude of the magnetic field in the cloud region. 
    The applied bias is $\Delta\phi=\SI{30}{\kilo\volt}$ and the RNG pressure is $p_n=\SI{1e-1}{\milli\bar}$. For both figures the squares represent numerical results extracted from the PIC simulations. The dashed-dotted line is a prediction using the model of section~\ref{sec:AnModel} but using only the external electric field for calculating the collision cross-sections while the circle dotted line is obtained using the full electric field (external plus self-consistent) extracted from the PIC simulations.}
    \label{fig:B_scan}
\end{figure}

\begin{figure}
    \centering
    \includegraphics[width=0.45\textwidth]{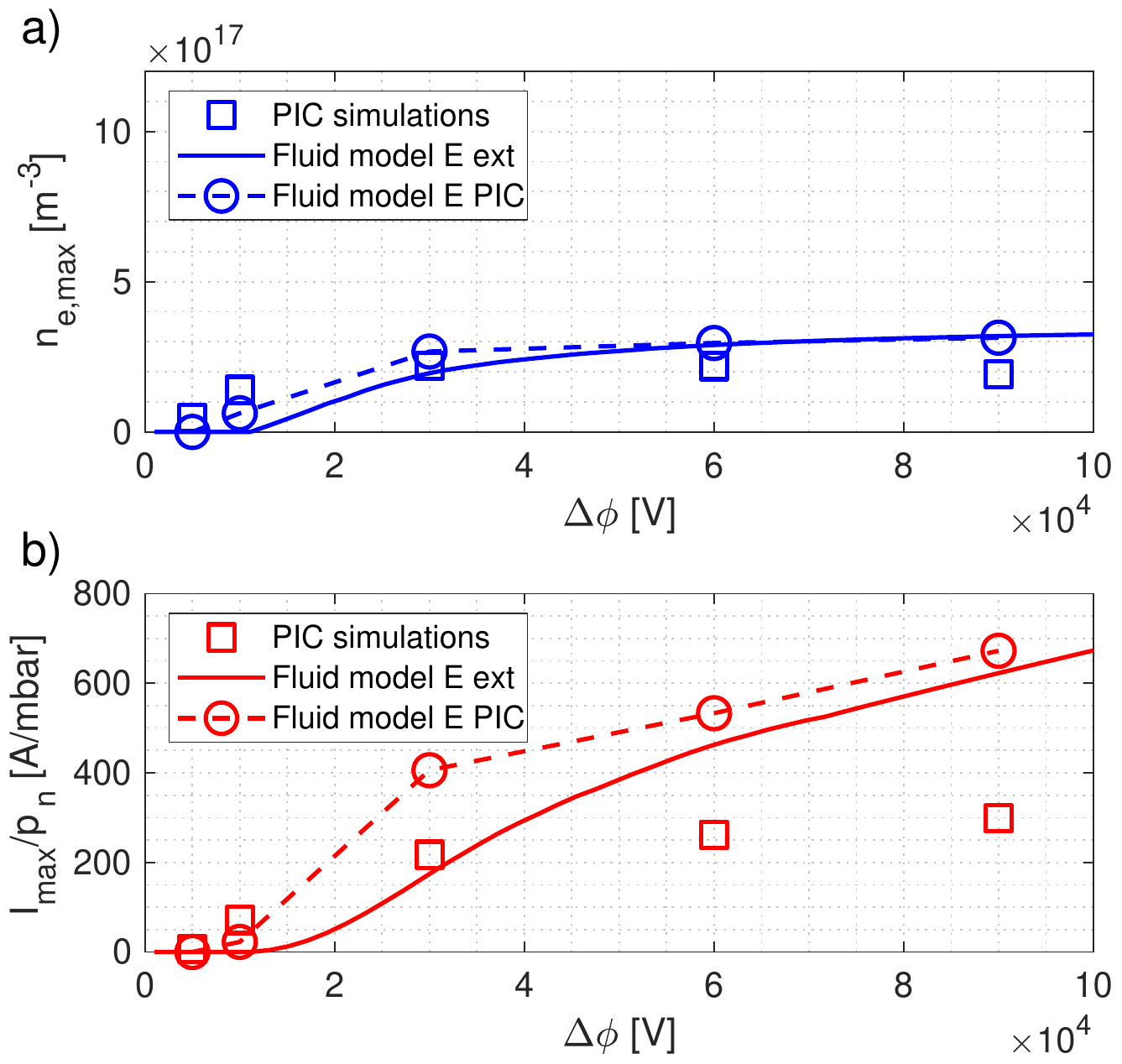}
    \caption{Same as Fig.~\ref{fig:B_scan} but as a function of the electric bias. The magnetic field amplitude is set at $B_{max}=\SI{0.28}{\tesla}$ and the RNG pressure is $p_n=\SI{1e-1}{\milli\bar}$.}
    \label{fig:E_scan}
\end{figure}

\begin{figure}
    \centering
    \includegraphics[width=0.45\textwidth]{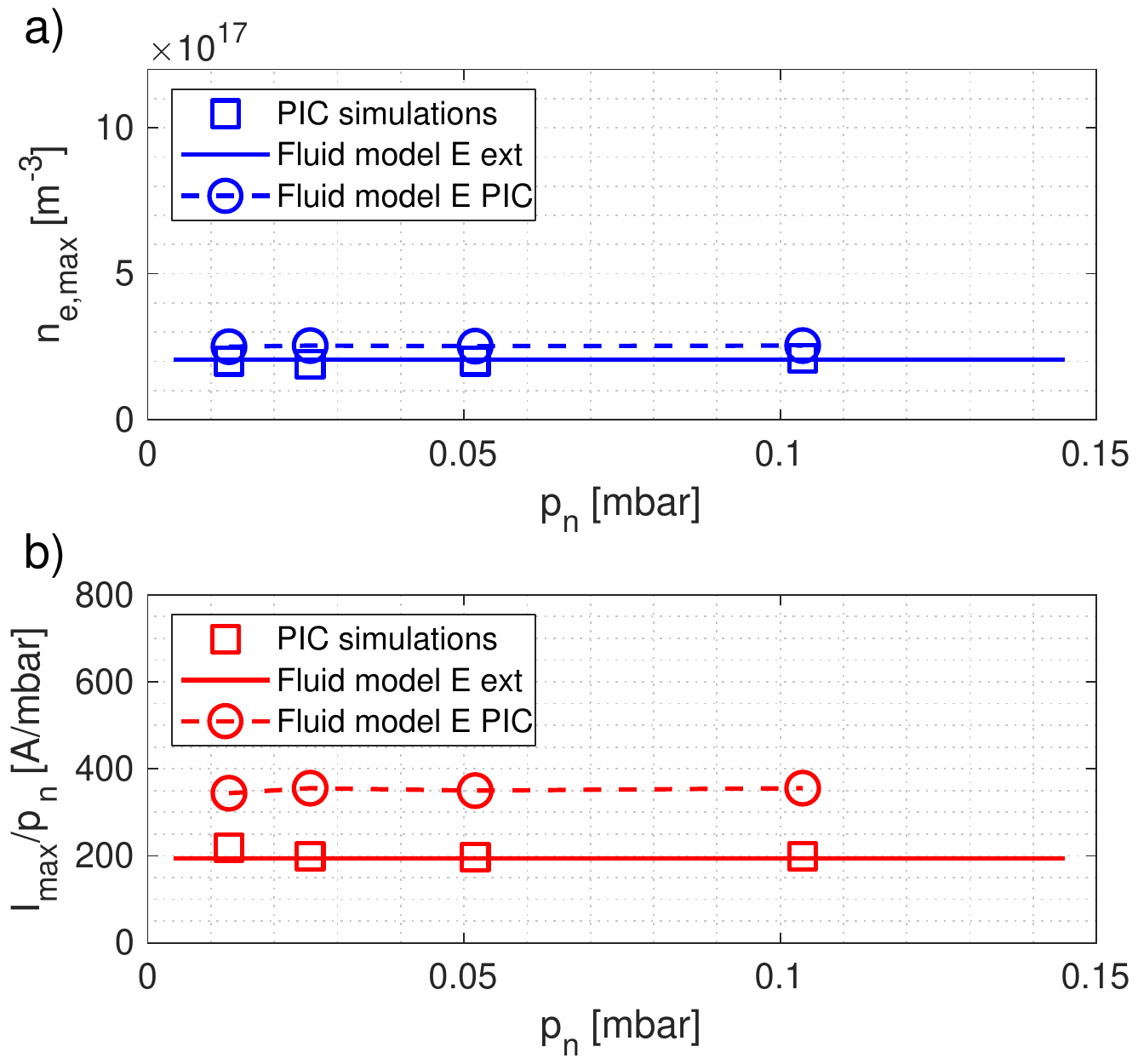}
    \caption{Same as Fig.~\ref{fig:B_scan} but as a function of the RNG pressure. The externally applied bias is set at $\Delta\phi=\SI{30}{\kilo\volt}$ and the magnetic field amplitude is set at $B_{max}=\SI{0.28}{\tesla}$.}
    \label{fig:nn_scan}
\end{figure}

To understand what are the operational parameters determining the peak density and current amplitudes, parametric scans are performed. The scanned parameters are the maximum magnetic field amplitude $B_{max}=0.14-\SI{0.56}{\tesla}$, the applied external bias $\Delta\phi=5-\SI{90}{\kilo\volt}$, and the RNG density $p_n=10^{-2}-10^{-1}\SI{}{\milli\bar}$. The results are shown respectively in Fig.~\ref{fig:B_scan}, Fig.~\ref{fig:E_scan}, and Fig.~\ref{fig:nn_scan}. We find that the peak density has a quadratic dependence on the magnetic field amplitude, see Fig.~\ref{fig:B_scan}(a), it has a non-trivial dependence on the external bias, see Fig.~\ref{fig:E_scan}(a), and it is independent of the RNG pressure, see Fig.~\ref{fig:nn_scan}(a). Based on the PIC simulation results, the radial current appears to scale linearly with $B_{max}$, see Fig.~\ref{fig:B_scan}(b), to have a non-trivial dependency on the external bias, see Fig.~\ref{fig:E_scan}(b), and finally to be linearly proportional to the RNG pressure, see Fig.~\ref{fig:nn_scan}(b). The external bias scan (Fig.~\ref{fig:E_scan}) suggests the existence of two regimes in the electron peak density and peak current depending on the externally applied bias. Above a certain bias, $\Delta\phi>\SI{30}{\kilo\volt}$, the slope of the curves changes drastically. An explanation is proposed for these dependencies in section~\ref{sec:AnModel}.

\subsection{Prevalent force terms}
\begin{figure}
    \centering
    \includegraphics[width=0.5\textwidth]{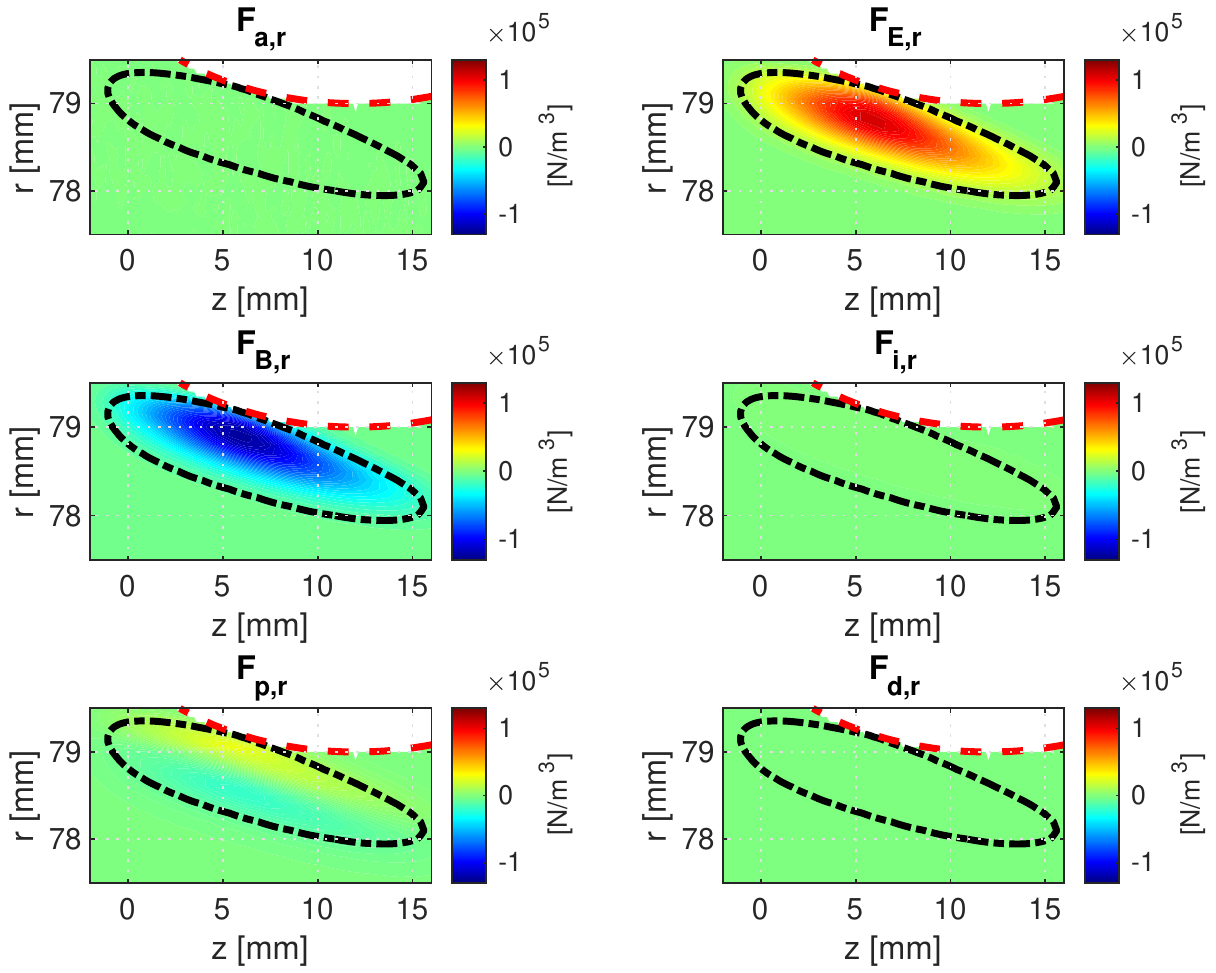}
    \caption{Contour plot of the different force terms in the radial fluid force balance equation at a time when the number of trapped particles is maximum. The black dashed-dotted line represents the cloud edge defined as the positions where $n=0.2\;n_{e,max}$. The red dashed line shows the metallic boundary. Here the subscript $_{r}$ denotes the projection along the radial direction.}
    \label{fig:Fr}
\end{figure}

\begin{figure}
    \centering
    \includegraphics[width=0.5\textwidth]{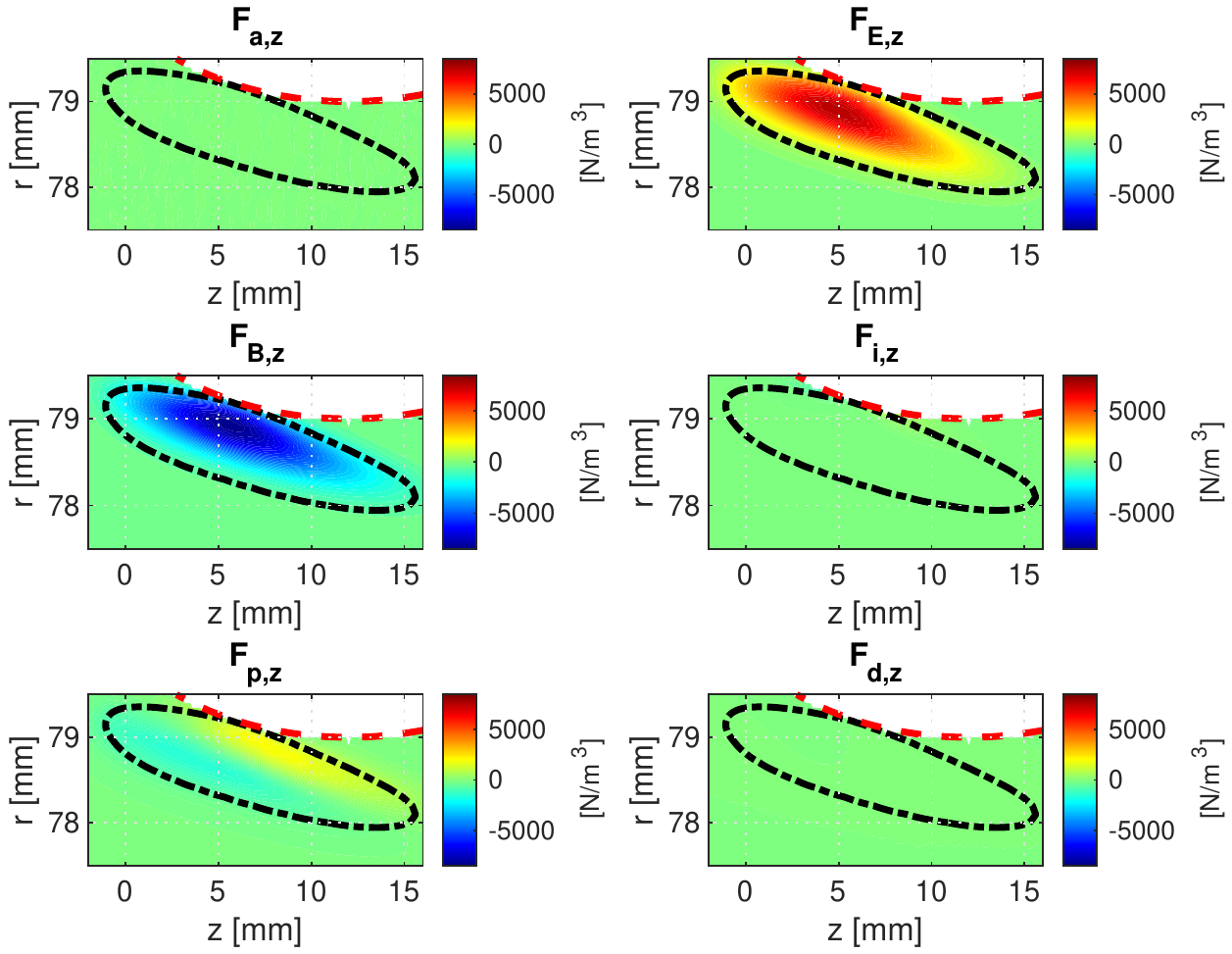}
    \caption{Same as Fig.~\ref{fig:Fr} for the axial direction. Here the subscript $_{z}$ denotes the projection along the axial direction.}
    \label{fig:Fz}
\end{figure}

\begin{figure}
    \centering
    \includegraphics[width=0.5\textwidth]{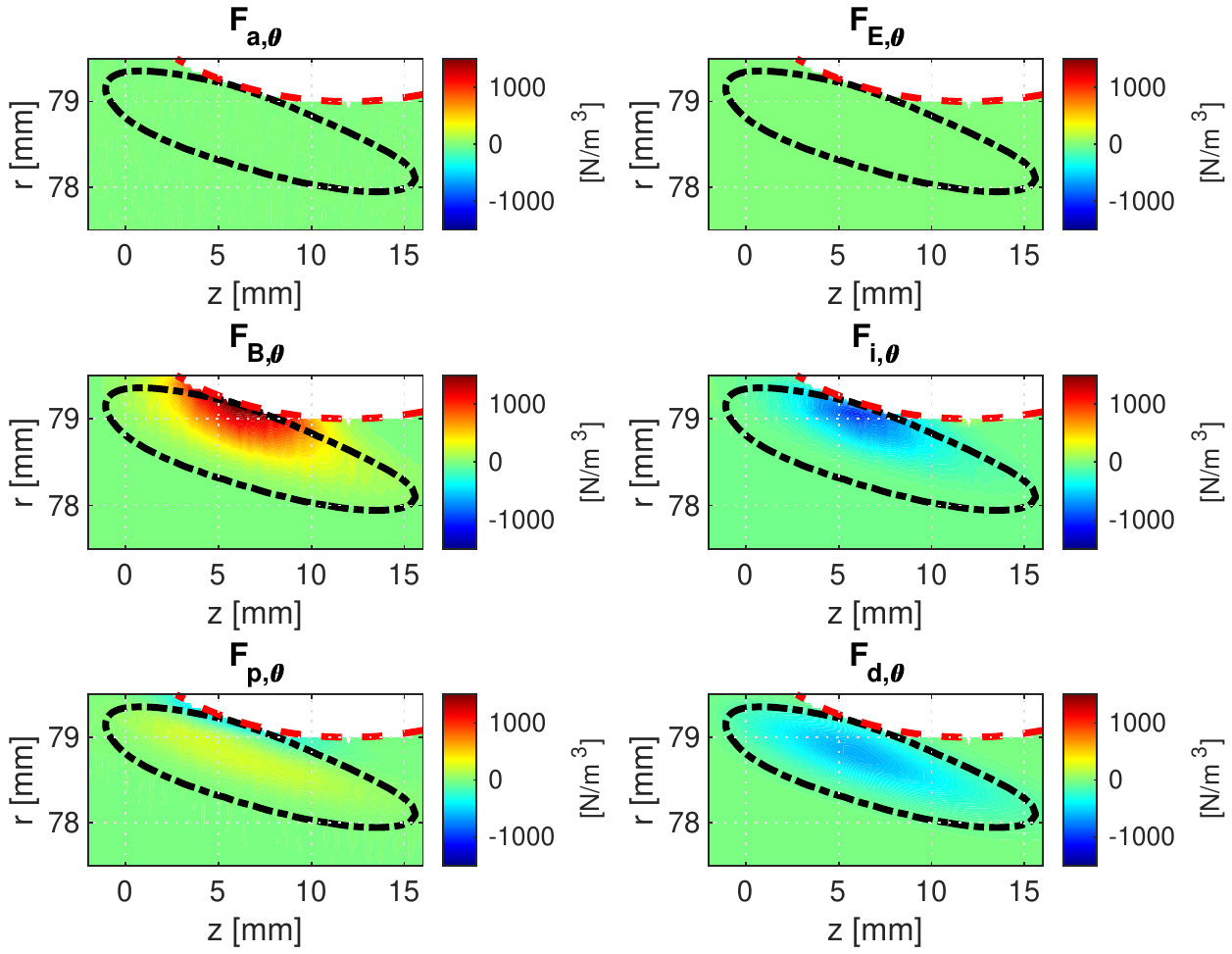}
    \caption{Same as Fig.\ref{fig:Fr} for the azimuthal direction. Here the subscript $_{\theta}$ denotes the projection along the azimuthal direction.}
    \label{fig:Fthet}
\end{figure}

To derive an analytical model that explains the parametric dependencies observed in the simulations, we evaluate the amplitude of the different forces acting on the electron fluid, using simulation results. These forces are evaluated by calculating the moments of the distribution function as extracted from the PIC simulations, at a time when the electron density is maximum, and are then averaged over several electron cyclotron periods to reduce numerical noise. The terms considered in the fluid model are: the electric and magnetic forces $\vec{F}_E=qn\vec{E}$ and $\vec{F}_B=qn\vec{u}\cross\vec{B}$; the inertial term $\vec{F}_i=-mn(\vec{u}\cdot\grad)\vec{u}$; the pressure term $\vec{F}_p=-\div\stackrel{\leftrightarrow}{P}$; the fluid acceleration term $\vec{F}_a=mn\partial_t\vec{u}$; the total collisional drag force term $\vec{F}_d=-n m n_n<\sigma_d v>_f\vec{u}$ which takes into account the effect of elastic collisions, the effect of ionisation collisions on the ionising electrons, and the effective ionisation drag due to the production of new low energy electrons by ionisation. Here, $m$ and $q$ are the electron mass and charge; $n$ the fluid density; $\vec{E}$ the total electric field taking into account external and self-generated components; $\vec{u}$ the fluid velocity; $\vec{B}$ the external magnetic field; $\stackrel{\leftrightarrow}{P}$ the pressure tensor; $<>_f$ denotes the average over the electron velocity distribution function; $v$ is the magnitude of the electron velocity, i.e. the electron speed; $\sigma_d$ is the total electron-neutral collision cross-section for momentum exchange; $n_n$ is the RNG density. In the radial and axial direction, as represented in Fig.~\ref{fig:Fr} and Fig.~\ref{fig:Fz} respectively, it can be observed that the dominant terms are the electric and magnetic forces, and that the pressure term is one order of magnitude smaller. In these directions, the inertial force term, and collisional drag term, are completely negligible. In the azimuthal direction, represented in Fig.~\ref{fig:Fthet}, the dominant terms are the inertial term, the magnetic force and the collisional drag. In this direction, the pressure force is also smaller than all the other terms. It is also relevant to observe that the sum of all the force terms in each direction is approximately 0 ($\vec{F}_i+\vec{F}_E+\vec{F}_B+\vec{F}_p+\vec{F}_d-\vec{F}_a\approx0$), as this is a strong verification for the implementation of the numerical model. 

\section{Analytical model}
\label{sec:AnModel}
In this section, an explanation for the parametric dependencies obtained from the simulations, see Fig.~\ref{fig:B_scan}, Fig.~\ref{fig:E_scan}, and Fig.~\ref{fig:nn_scan}, is proposed.

\subsection{Fluid-Poisson Model}
A prediction for the cloud average density and the average radial current density can be derived by considering the electron fluid equations coupled to Poisson's equation and neglecting electron pressure effects, which is a simplification that is quantitatively justified by our PIC simulations. We start from the fluid force balance equation:
\begin{equation}
    m n \left(\pdv{\vec{u}}{t} + (\vec{u}\cdot\grad)\vec{u}\right)=n q(\vec{E}+\vec{u}\cross\vec{B}) + \vec{F}_d.
    \label{eq:Ftot}
\end{equation}
As shown in Fig.~\ref{fig:Fr}, the dominant radial forces in the bulk of the cloud are the electric and magnetic forces, thus the radial component of Eq.~\eqref{eq:Ftot} gives:
\begin{equation}
    u_\theta=-\frac{E_r}{B_z}.
    \label{eq:uthet}
\end{equation}
As can be seen in Fig.~\ref{fig:Fthet}, the dominant forces in the azimuthal direction are the inertial force ($\vec{u}\cdot\grad\vec{u}$ term), the magnetic force, and the drag force. Since $B_r\ll B_z$ and in the cloud we also have $u_z\ll u_r$, we can assume $u_zB_r\ll u_rB_z$ and, hence, the azimuthal component of Eq.~\eqref{eq:Ftot} gives, at equilibrium:
\begin{equation}
    m u_r \frac{1}{r}\pdv{r}(r u_\theta)=-q u_r B_z - m n_n<\sigma_d v>_f u_\theta.
    \label{eq:Fr}
\end{equation}
The left hand side of Eq.~\eqref{eq:Fr} can be rewritten by using the expression for the azimuthal velocity, Eq.~\eqref{eq:uthet}, using Gauss's law, and assuming $|\partial_z E_z|\ll |\frac{1}{r} \partial_r(rE_r)|$ and $\frac{1}{r} \partial_r(r/B_z)\ll\frac{1}{r} \partial_r(rE_r) $. We obtain:
\begin{equation}
    m u_r \frac{1}{r}\pdv{r}(r u_\theta)= -m \frac{u_r}{B_z} \frac{1}{r}\pdv{r}r E_r=-m \frac{u_r}{B_z}\frac{q n}{\epsilon_0}.
    \label{eq:assumpt}
\end{equation}
Rewriting Eq.~\eqref{eq:Fr} then gives the radial fluid velocity as:
\begin{equation}
    u_r= -\frac{q n_n <\sigma_d v>_f}{m}\frac{E_r}{\omega_{p}^2-\Omega_{c}^2}.
    \label{eq:ur}
\end{equation}
Here, $\Omega_{c}=qB/m$ is the cyclotron frequency and $\omega_{p}=\sqrt{q^2n/(\epsilon_0 m)}$ is the plasma frequency.

The time-averaged density can be obtained by starting from the time average of the continuity equation:
\begin{equation}
    \left<\pdv{n}{t}+\div(n \vec{u})\right>_T=\left<n n_n <\sigma_{io} v>_f\right>_T,
\end{equation}
where $\sigma_{io}$ is the ionisation cross-section and $<>_T$ denotes the time average over one cloud breathing oscillation. Considering the case of density oscillations at the spatial peak density: 
\begin{equation}
    \left<\pdv{n}{t}\right>_T=0\; \text{and}\; \grad n =0;
\end{equation}
assuming dominant radial losses, azimuthal symmetry and using the radial velocity obtained in Eq.~\eqref{eq:ur}, the continuity equation can be rewritten as:
\begin{equation}
    -\frac{q}{m} \left<\frac{n}{r}\pdv{r}r\left[\frac{ n_n <\sigma_d v>_f}{\omega_{p}^2-\Omega_{c}^2} E_r\right]\right>_T = \left<n n_n <\sigma_{io} v>_f\right>_T.
\end{equation}
Using Gauss's law and the fact that $\grad n =0$ at the peak density once more, as well as the assumptions used in Eq.~\eqref{eq:assumpt}, we obtain an expression for the time averaged plasma frequency at the spatial peak:
\begin{equation}
    \omega_{p,max}^2= \Omega_{c}^2\left<\frac{<\sigma_{io}v>_f}{<\sigma_{io}v>_f+<\sigma_d\ v>_f}\right>_T.
    \label{eq:omegape_pred}
\end{equation}
which gives directly the average cloud density at the spatial peak $n_{e,max}$.

Using the radial velocity and average density previously derived, and assuming zero axial velocity, we can also obtain an estimate for the peak current by integrating the loss term $\div n\vec{u} $ over the cloud volume:
\begin{equation}
    I=\int <q \div (n \vec{u})>_T dV \approx -2\pi L r_+ \epsilon_0 n_n <E_r <\sigma_{io} v>_f>_T.
    \label{eq:current}
\end{equation}
Where $L$ is the characteristic cloud axial length and $r_+$ the cloud outer radial limit. These geometric quantities can be estimated from the potential well dimensions in vacuum. It can be observed that this current is linearly proportional to the RNG density, but has a more complex scaling in electric and magnetic field due to the non-trivial dependency of $<\sigma_{io}v>$ on these terms.

\subsection{Model verification}

\begin{figure}
    \centering
    \includegraphics[width=0.48\textwidth]{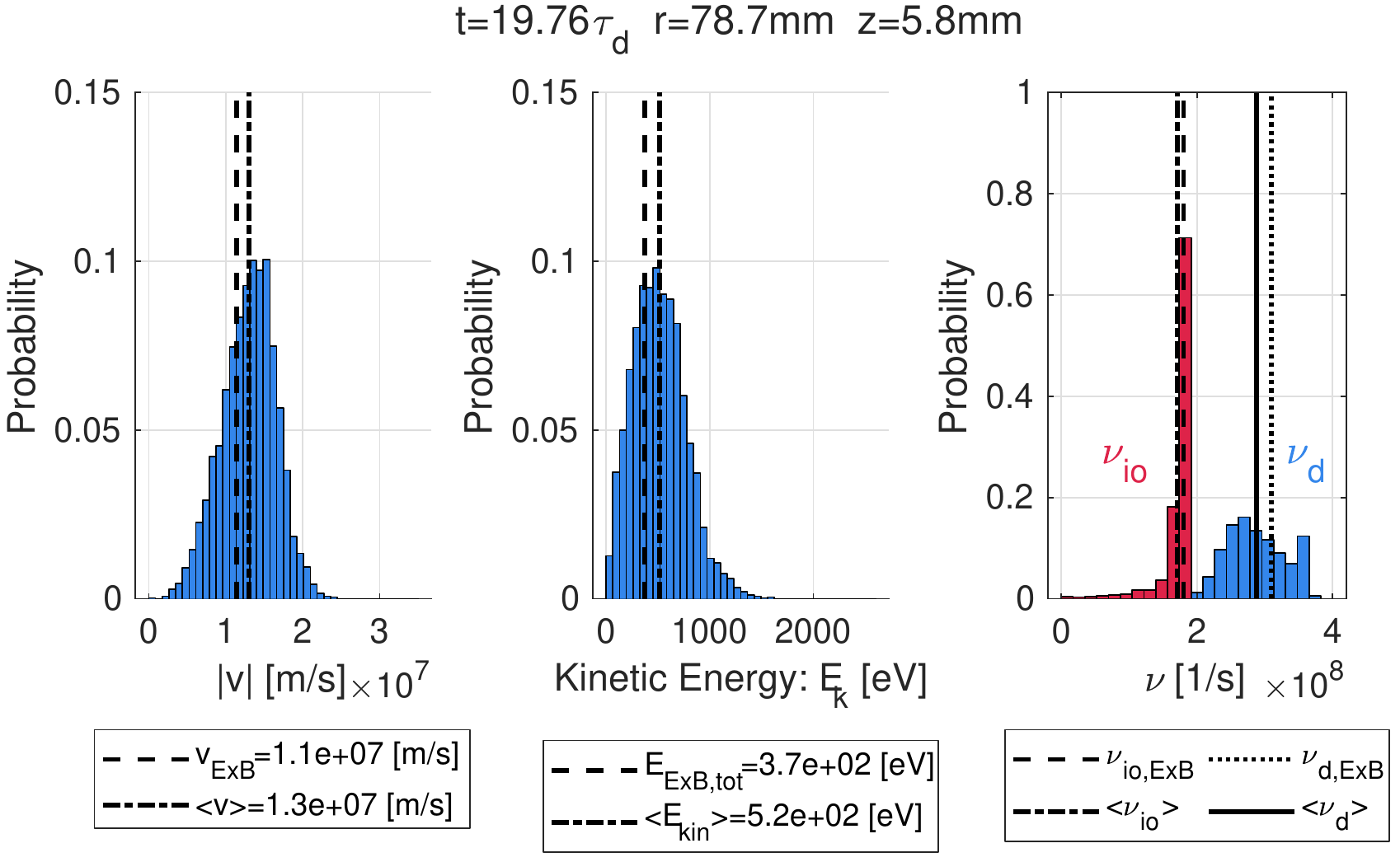}
    \caption{Electron velocity distribution (left), kinetic energy distribution (middle), and collision frequencies distribution (right), extracted from the PIC simulation shown in Fig.~\ref{fig:tes30kv_ince2Fields} and Fig.~\ref{fig:tes30kv_timeevol}, and represented at a time when the density is maximum, and at the position of peak density. For comparison, the average is represented by the dash-dotted and solid lines. The dashed lines and the dotted line is the equivalent quantity if the velocity of the electron is exactly the local $\vec{E}\cross\vec{B}$ velocity.}
    \label{fig:Vdistrib}
\end{figure}

To predict the peak electron density and current using Eq.~\eqref{eq:omegape_pred} and Eq.~\eqref{eq:current}, it is necessary to calculate the collision frequencies by averaging $\sigma v$ over the electron velocity distribution function. As can be observed in Fig.~\ref{fig:Vdistrib}, due to the large $\vec{E}\cross\vec{B}$ drift, the average speed of electrons is dominated by the azimuthal component of the average velocity such that $\vec{u}\approx\vec{E}\cross\vec{B}/B^2$. Furthermore, the variance of the speed distribution and the variance of the kinetic energy distribution are relatively small. We may therefore approximate the collision frequencies $\nu$ as follows:
\begin{equation}
    \nu=n_n<v\sigma(E_k)>_f\approx n_n\left|\frac{\vec{E}\cross\vec{B}}{B^2}\right|\sigma\left(\frac{1}{2}m\frac{|\vec{E}\cross\vec{B}|^2}{B^4}\right).
    \label{eq:nuapprox}
\end{equation}
Here, $E_k=mv^2/2$ is the kinetic energy of the electrons. The quality of this approximation can be assessed by looking at the figure on the right-hand side of Fig.~\ref{fig:Vdistrib}, where we see a close agreement between the collision frequencies directly computed from the PIC simulations, and the approximate collision frequencies given by Eq.~\eqref{eq:nuapprox}. With these definitions, Eq.~\eqref{eq:omegape_pred} and Eq.~\eqref{eq:current} are used to verify the analytic model for the parametric scans of section~\ref{sec:results}. To calculate the collision frequencies, the drift velocity is calculated once using only the externally imposed electric field (dashed line in Figs.~\ref{fig:B_scan},~\ref{fig:E_scan},~\ref{fig:nn_scan}), and once using the total electric field extracted from the simulations (circles in Figs.~\ref{fig:B_scan},~\ref{fig:E_scan},~\ref{fig:nn_scan}). The comparison between the simulation results and the model predictions show that the knowledge of the total electric field is necessary to obtain the correct scaling. When using the total electric field in the model, Eq.~\eqref{eq:omegape_pred} and Eq.~\eqref{eq:current}, the analytical scalings for the scans on $B_{max}$, see Fig.~\ref{fig:B_scan}, on $\Delta\phi$, see Fig.~\ref{fig:E_scan}, and on $p_n$, see Fig.~\ref{fig:nn_scan}, are well reproduced for $n_{e,max}$, and the trends are captured for $I_{max}$. The slightly worse agreement for the current could be explained by the approximated size of the cloud which is assumed independent of the external parameters. These results reveal that a model for the self-consistent electric field is needed to have an analytical prediction from the reduced model, and to limit the need for computationally expensive numerical simulations.

It appears from this reduced model that a fluid code could be sufficient to study the problem at hand. However, this model considers only the time-averaged behavior of the cloud, and it is not yet known if kinetic effects are important to describe the dynamics (cloud breathing). Moreover, in a fluid code, the implementation of boundary conditions for the fluid would be much more complicated. Furthermore, as seen in Fig.~\ref{fig:Fthet}, the amplitude of the pressure force in the azimuthal direction, while small, is not completely negligible. This means that, if this term is important to describe the dynamics of the system, the use of an isothermal fluid model is not justified for a two-dimensional fluid model, and a more complex closure equation is necessary.

\subsection{Collision regimes}

\begin{figure}
    \centering
    \includegraphics[width=0.48\textwidth]{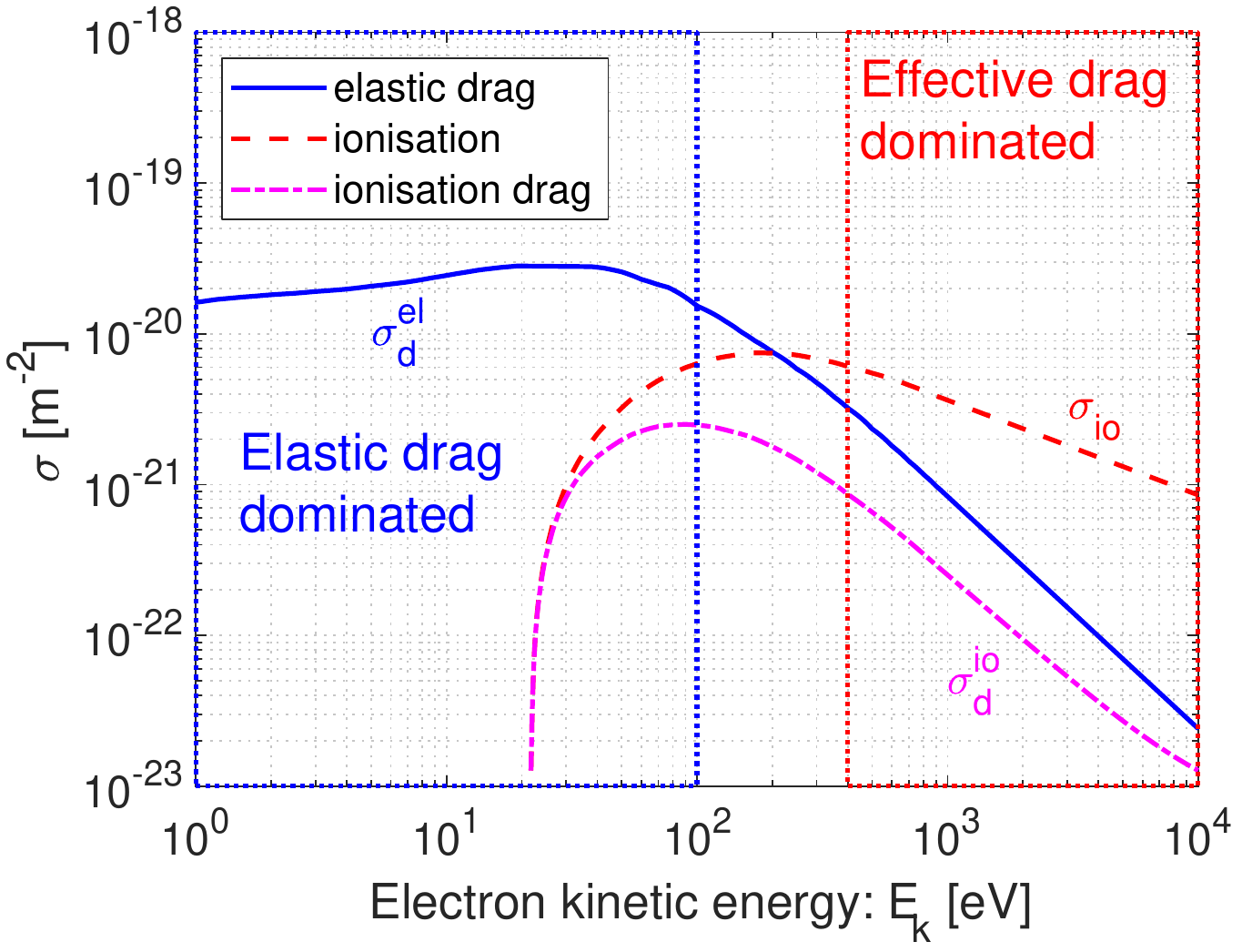}
    \caption{Electron collision cross-sections of \ce{Ne} atoms as a function of the electron kinetic energy. The neutral particles are assumed to have zero velocity. }
    \label{fig:sigma_compare}
\end{figure}

In the parametric scans on the external bias of Fig.~\ref{fig:E_scan}, two regimes have been identified for both the current and the peak electron density, where a plateau is reached at large biases (above $\Delta\phi=\SI{30}{\kilo\volt}$). This result can be explained by examining the collision cross-sections, represented in Fig.~\ref{fig:sigma_compare}, and their corresponding effective drag frequencies. Here, two main regimes can be defined. For low electron kinetic energies, $E_k\lesssim\SI{100}{\electronvolt}$, the elastic drag dominates and the peak densities depend directly on the electron energies and by extension on the externally applied electric and magnetic fields, see Eq.~\eqref{eq:nuapprox} and Eq.~\eqref{eq:omegape_pred}. On the contrary, for high electron kinetic energies, $E_k\gtrsim\SI{400}{\electronvolt}$, the effective drag due to electron creation dominates. This means that in Eq.~\eqref{eq:omegape_pred}, $\sigma_d=\sigma_{io}+\sigma_d^{el}+\sigma_d^{io}\approx\sigma_{io}$ and the peak electron density becomes independent of the electron energies.
\begin{equation}
\begin{split}
    \omega_{pe,peak}^2&= \left<\Omega_{ce}^2\frac{<\sigma_{io}v>_f}{<\sigma_{io}v>_f+<\sigma_d v>_f}\right>_T\approx\frac{1}{2}\Omega_{ce}^2.
\end{split}
\end{equation}
In the PIC simulations, this change of regime is expected for biases $\Delta\phi\approx\SI{30}{\kilo\volt}$ where the electron kinetic energy is $E_k\approx\SI{200}{\electronvolt}$. Furthermore, as the current is linearly proportional to the cloud density, two regimes are also expected in the dependence on the external bias for this quantity.

\section{Discussion and Outlook}
\label{sec:discussion}
A new PIC code has been developed that is able to simulate the self-consistent formation of an electron cloud in a configuration close to the one existing in gyrotron electron guns. It is shown that electron clouds form inside the potential well (which results from a combination of the external magnetic field and of the self-consistent electric field) and that such clouds are the source of important radial currents. Parametric scans are presented that show the dependency of the peak electron cloud density and radial current on external parameters. To explain the observed dependencies, an analytical fluid-Poisson model is derived and successfully validated with the simulation results using the PIC code. This model shows that the radial current is linearly dependent on the RNG pressure, which permits the artificial increase of the collision frequencies, and is important for the applicability of the code and its relevance in the design process of gyrotron guns (to provide shorter calculation times). This model also confirms the quadratic dependence of the peak density on the magnetic field amplitude. Finally, this article reveals and explains that, depending on the trapped electrons kinetic energies, two collision regimes can be identified. In the first regime, the radial transport is dominated by elastic collision drag and the electron cloud density is proportional to the electric and magnetic field amplitude. In contrast, in the second regime the transport is dominated by the effective drag imposed by the release of electrons from ionisation, and the electron density is only proportional to the magnetic field amplitude.
It is however important to stress that this model does not capture the dynamics of the system and cannot explain the oscillations in the radial current and peak density. The development of an extended model relaxing the equilibrium assumption is planned, and could encompass pressure effects, density gradient effects, or non uniform energy distributions effects currently neglected in the reduced model. Furthermore, the radial currents, when scaled using a realistic pressure of $p_n=\SI{1e-8}{\milli\bar}$ are of the order of $I\approx\SI{1}{\micro\ampere}$ which is too low compared to previous experiments, and would not cause detrimental effects. These discrepancies can potentially be explained by a rapid loss of particles due to diocotron instabilities, which may lead to current pulses, or by an oversimplification of the geometry and of the resulting potential well. To verify these hypotheses, one solution would be to break the azimuthal symmetry, which is, however, numerically expensive. Another solution would be to use models based on diocotron linear stability equations with density profiles extracted from the current PIC code. To study geometric effects, the current Poisson solver is currently being adapted to allow for more realistic geometries. 
In parallel to this study, and to the numerical research, a new experiment is currently being developed at the Swiss Plasma Center. Such experiment will reproduce the key characteristics of the gyrotron electron gun in particular in terms of geometries, applied fields amplitudes and topologies, and pressure. The aim is to study such electron clouds. This will broaden the view on the current subject and provide input to address the weak points of the presented model. The experiment will further guide the theoretical research and help validating the current PIC code and reduced models.   

\begin{acknowledgments}
A. Cerfon wishes to thank Paolo Ricci and the Swiss Plasma Center at EPFL for hosting him during the academic year 2020-2021, during which some of the research work presented here was performed. The authors also wish to thank Trach Minh Tran, Patryk Kaminski and Jérémy Genoud, for the development of FEM libraries and the initial version of the PIC code on which the current code is built.

This work has been carried out within the framework of the EUROfusion Consortium, funded by the European Union via the Euratom Research and Training Programme (Grant Agreement No~101052200 — EUROfusion). Views and opinions expressed are however those of the author(s) only and do not necessarily reflect those of the European Union or the European Commission. Neither the European Union nor the European Commission can be held responsible for them.

The calculations have been performed using the facilities of the Scientific IT and Application Support Center of EPFL.

This work was supported in part by the Swiss National Science Foundation under grant No.~204631.
\end{acknowledgments}
\section*{AUTHOR DECLARATIONS}
\subsection*{Conflict of Interest}
The authors have no conflicts to disclose.
\section*{DATA AVAILABILITY}
The data that support the findings of this study are
available from the corresponding author upon reasonable
request.

\bibliographystyle{ieeetr}
\bibliography{bibliography}

\end{document}